\documentclass[10pt, twocolumn, conference]{IEEEtran}
\IEEEoverridecommandlockouts

\usepackage{subfigure}
\usepackage{paralist}
\usepackage{color}
\usepackage{url}
\usepackage[linesnumbered,ruled,boxed,vlined]{algorithm2e}

\usepackage{amsmath}
\usepackage{amssymb}
\usepackage{graphicx}
\usepackage{bm}
\usepackage{paralist}
\usepackage{subfigure}
\usepackage{times}
\usepackage{balance}
\usepackage{cite}
\usepackage{comment}

\newcommand{\separator}{
  \begin{center}
    \rule{\columnwidth}{0.3mm}
  \end{center}
}

\def\ie{{\it i.e.}}

\newtheorem{theorem}{Theorem}

\newtheorem{proposition}{Proposition}
\newtheorem{lemma}{Lemma}
\newtheorem{definition}{Definition}

\newcommand{\prob}[1]{\mathbb{P}[ #1 ]}

\newcommand{\beq}{\begin{eqnarray*}}
\newcommand{\eeq}{\end{eqnarray*}}
\newcommand{\beqn}{\begin{eqnarray}}
\newcommand{\eeqn}{\end{eqnarray}}
\newcommand{\bemn}{\begin{multiline}}
\newcommand{\eemn}{\end{multiline}}

\newcommand{\note}[1]{{\color{red}{[\textit{#1}}]}}

\allowdisplaybreaks
\begin{document}

\title{Necessary and Sufficient Budgets in Information Source Finding with Querying: Adaptivity Gap 
              }

\author
{Jaeyoung Choi and Yung Yi$^\dag$
\thanks{ $\dag$: Department of EE, KAIST,
Republic of Korea. (e-mails: \{jychoi14,yiyung\}@kaist.ac.kr }
}

\maketitle

\begin{abstract}
In this paper, we study a problem of detecting the source of diffused information by querying individuals, given a sample snapshot of the information diffusion graph, where two queries are asked:
{\em (i)} whether the respondent is the source or not, and {\em (ii)} if not, which neighbor spreads the information to the respondent.
We consider the case when respondents may not always be truthful and some cost is taken for each query.
Our goal is to quantify the necessary and sufficient budgets to achieve the detection probability $1-\delta$ for any given $0<\delta<1.$ To this end, we study
two types of algorithms: adaptive and non-adaptive ones, each of which corresponds to whether
we adaptively select the next respondents based on the
answers of the previous respondents or not.
We first provide the information theoretic lower bounds for the necessary budgets in both algorithm types.
In terms of the sufficient budgets, we propose two practical estimation algorithms, each of non-adaptive and adaptive types, and for each algorithm,
we quantitatively analyze the budget which ensures $1-\delta$ detection accuracy. This theoretical analysis not only quantifies the budgets needed by practical estimation algorithms achieving a given target detection accuracy in finding the diffusion source, but also
enables us to quantitatively characterize the amount of extra budget required in non-adaptive type of estimation, refereed to as {\em adaptivity gap}.
We validate our theoretical findings over synthetic and real-world social network topologies.
\end{abstract}



\section{Introduction}

Information diffusion in networks can be used to model many real-world
phenomena such as propagation of infectious diseases, diffusion of
a new technology, computer virus/spam infection in the Internet,
and tweeting and retweeting of popular topics. A problem of finding the information source is to
  identify the true source of the information spread. This is clearly
  of practical importance, because harmful diffusion
  can be mitigated or even blocked, e.g., by vaccinating humans or installing
  security updates \cite{Kai2016}.
  Recently, extensive research attentions for this problem
have been paid for various network topologies and diffusion models
\cite{shah2010,shah2012,zhu2013,Luo2013,bubeck2014,Kai2016, Chang2015,farajtabar2015}, whose major
interests lie in constructing an efficient estimator and providing
theoretical analysis on its detection performance.

Prior work directly or indirectly conclude that this information source finding turns out to be a challenging task unless sufficient side information
or multiple diffusion snapshots are provided.
There have been several research efforts which use multiple snapshots \cite{Zhang2014} or a side information about a restricted superset the true source belongs \cite{dong2013}, thereby the detection performance is significantly improved. Another type of side information is the one obtained from {\em querying}, \ie, asking questions to a subset of infected nodes and gathering more hints about who would be the true information source \cite{Choi17}. The focus of this paper is also on querying-based approach (we will shortly present the difference of this paper from \cite{Choi17} at the end of this section).

In this paper, we consider an \emph{identity with direction} (id/dir in short) question as follows. First, a querier asks an identity question of
whether the queriee is the source or not, and if "no", the queriee is subsequently asked the direction question of  which neighbor
spreads the to the queriee. Queriees may be untruthful with some
probability so that the multiple questions to the same queriee are allowed to filter the untruthful answers, and the total number of questions can be asked within a given budget.
We consider two types of querying schemes: {\em (a) Non-Adaptive ({\bf NA})} and {\em (b) ADaptive ({\bf AD}).} In {\bf NA}-querying,
    a candidate queriee set is first chosen,
    and the id/dir queries are asked in a batch manner.
    In {\bf AD}-querying, we start with some
initial quieree, iteratively ask a series of id/dir questions to the current queriee, and adaptively determine the next queriee using the
(possibly untruthful) answers from the previous queriee, where this iterative
querying process lasts until the entire budget is used up.

We summarize our main contributions of this paper.
First, we obtain the necessary budgets for both querying schemes to achieve the $(1-\delta)$ detection probability for any given $0<\delta<1.$
To this end, we establish information theoretical lower bounds
from the given diffusion snapshot and the answer samples from querying.
Our results show that it is necessary to use the budget $\Omega \left(\frac{(1/\delta)^{1/2}}{\log(\log(1/\delta))}\right)$
for the {\bf NA}-querying whereas $\Omega \left(\frac{\log^{1/2} (1/\delta)}{\log(\log(1/\delta))}\right)$
for the {\bf AD}-querying, respectively.
Second, to obtain the sufficient amount of budget for $(1-\delta)$ detection performance, we consider two estimation algorithms, each for both querying schemes, based on a simple majority voting to handle the untruthful answer samples. We analyze simple, yet powerful estimation algorithms and
  characterize their detection probabilities for given parameters.
  Our results show that
it suffices to use $O \left(\frac{(1/\delta)}{\log(\log(1/\delta))}\right)$
for the {\bf NA}-querying, whereas
$O \left(\frac{\log^2 (1/\delta)}{\log(\log(1/\delta))}\right)$ is sufficient for the {\bf AD}-querying, respectively. The gap between necessary and sufficient budgets in both querying schemes is due to our consideration of simple, yet practical estimation algorithms based on majority voting, caused by the fact that the classical ML-based estimation is computationally prohibitive and even its analytical challenge is significant.
Our quantification of necessary and sufficient budgets enables us to
obtain the lower and upper bounds of the \emph{adaptive gap}, i.e., the gain of adaptive querying scheme compared to non-adaptive one.
Finally, we validate our findings via extensive simulations over popular random graphs (\emph{Erd\"{o}s-R\'{e}nyi} and scale-free graphs) and a real-world
  Facebook graph.

We end this section by presenting the difference of this paper from our preliminary work \cite{Choi17}. In \cite{Choi17},
(i) only identity question in the non-adaptive case is considered and
(ii) untruthfulness for the answers of identity questions in the adaptive case is not modeled. In this paper, we generalize and complete the model in terms of query types and schemes, which add non-negligible analytical challenges, and we establish information-theoretic lower bounds for the necessary amount of budget, which is the key step to quantifying the adaptivity gap.

\section{Model Preliminaries}
\label{sec:model}
\subsection{Diffusion Model and MLE}

We consider an undirected graph
$G=(V,E),$ where $V$ is a countably infinite set of nodes and $E$ is
the set of edges of the form $(i,j)$ for $i, j\in V$.  Each node
represents an individual in human social networks or a computer host
in the Internet, and each edge corresponds to a social relationship
between two individuals or a physical connection between two Internet
hosts. As an information spreading model, we consider a \emph{Susceptible-Infected} (SI) model 
under exponential distribution with rate of $\lambda_{ij}$ for the edge $(i,j),$ and all nodes are initialized to be
susceptible except the information source. Once a node $i$ has an information,
it is able to spread the information to another node $j$ if and only if
there is an edge between them. We
denote by $v_1 \in V$ the information source, which acts as a node that
initiates diffusion and denote by $V_N \subset V$, $N$ infected nodes
under the observed snapshot $G_N\subset G$. 
In this paper, we consider the case when $G$ is a regular tree, the diffusion rate $\lambda_{ij}$ is homogeneous with unit rate, \ie, $ \lambda_{ij}=\lambda =1$, and $N$ is
large, as done in many prior work
\cite{shah2010,shah2012,Khim14,dong2013,Zhang2014}.
We assume that there is no prior distribution about the source, \ie,
the uniform distribution.
As a useful prior result, under the SI-diffusion with homogeneous
rate over regular tree, the authors \cite{shah2010} first show that
the source chosen by the Maximum Likelihood Estimator (MLE) becomes the node with a highest 
graph-theoretic score metric, called {\em rumor centrality}.  
Formally, the estimator chooses $v_{RC}$ as the rumor source defined as
$v_{RC} = \arg\max_{v \in V_N} \mathbb{P}(G_N | v=v_1)$ where
$v_{RC}$ is called \emph{rumor center} (RC).


\subsection{Querying Model and Algorithm Classes}


\noindent{\bf \em Querying with untruthful answers.}
Using the diffusion snapshot of the information, a detector performs querying which refers to a process of asking some questions. We assume that a fixed budget $K$
is given to the detector (or the querier) and a unit budget has worth of asking one pair of id/dir question, \ie, 
 ``Are you the source?'' first and if the respondent answers ``yes" then it is done. Otherwise, the detector subsequently asks a direction question as ``Which neighbor spreads the information to you?''.
  In answering a query, we consider that each respondent $v$ 
  is only partially truthful in answering id and dir questions, with probabilities of being truthful, $p_v$ and $q_v$, respectively. To handle untruthful answers, the querier may ask to a respondent $v$ the question multiple times, in which $v$'s truthfulness is assumed to be independent. We also assume that homogeneous truthfulness across individuals, \ie,  $p_v = p$ and $q_v = q$ for all $v \in V_N,$ and $p > 1/2, q>1/d$ meaning that all answers are more biased to the truth.  In terms of querying schemes, we consider the following two types, {\em non-adaptive} and {\em adaptive,} for each of which we restrict ourselves 
  to a certain class of querying mechanisms:

\begin{figure}
\subfigure[Non-adaptive (NA)-querying.]{\includegraphics[width=0.23\textwidth]{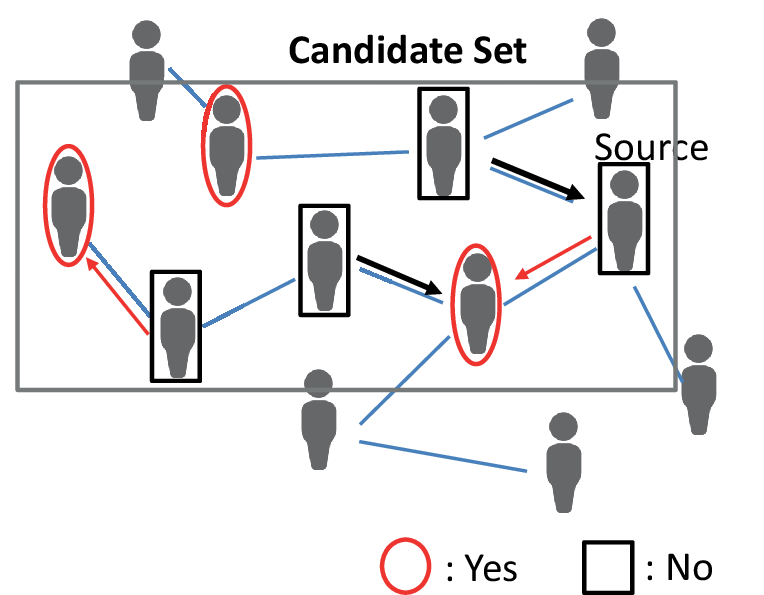}\label{fig:Noninter}}
\subfigure[Adaptive (AD)-querying.]{\includegraphics[width=0.23\textwidth]{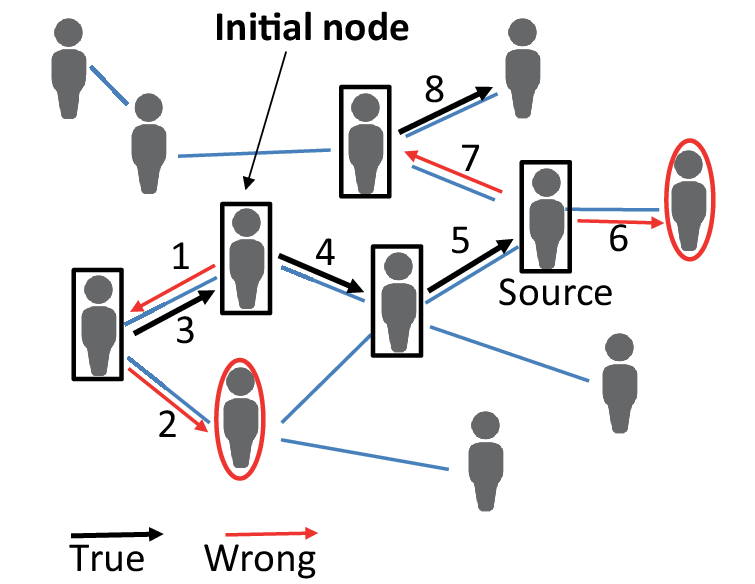}\label{fig:Inter}}
\hspace{0cm}
\caption{Examples of two querying types with untruthful
  answers ($r=1$). In (a), the querier selects a candidate set (a large square) and asks just one id/direction question in a batch manner under the untruthful answers. In (b),
  starting from the initial node, the querier first asks one id/direction question and adaptively tracks the true source with the untruthful answers. (In (b), True is the direction of true parent and Wrong is the wrong direction.)}
\label{fig:querying}
 \vspace{-0.2cm}
\end{figure}




\smallskip
\noindent {\bf \em NA-querying.} In this querying, we first choose a subset of infected nodes {\em in a batch} as a candidate set which is believed to contain the true source, then ask (multiple) id/dir question to each respondent inside the candidate set, and finally run an estimation algorithm based on the answers from all the respondents. We consider the following class of {\bf NA}-querying mechanisms, denoted by $\mathcal{NA}(r,K)$, in this paper:


\smallskip
\begin{definition}(Class $\mathcal{NA}(r,K)$)
\label{def:na}
In this class of {\bf NA}-querying schemes with the parameter $r$ 
and a given budget $K,$
the querier first chooses the candidate set of $\left\lfloor K/r \right\rfloor$ infected nodes according to the following selection rule: 
We initially select the node RC and 
add other infected nodes in the increasing sequence in terms of the {\em hop-distance} from the RC. Then, the querier asks the id/dir question $r$ times to each node in the selected candidate set. 
\end{definition}

\smallskip
\noindent {\bf \em AD-querying.}
A querier first chooses an initial node to ask the id/dir question, possibly multiple times, and 
  the querier adaptively determines the next respondent using the answers from the previous queriee, which is repeated until the entire budget is exhausted. We consider the following class of AD-querying mechanisms, denoted by $\mathcal{AD}(r,K)$, in this paper:


\smallskip
\begin{definition}(Class $\mathcal{AD}(r,K)$)
\label{def:ad}
In this class of {\bf AD}-querying schemes with the parameter $r$ and a given budget $K,$ 
the querier first chooses the RC as a starting node, and performs the repeated procedure mentioned earlier, but in choosing the next respondent, we only consider one of the neighbors of the previous node, where each chosen respondent is asked the id/dir question $r$ times. If the querier can not obtain any information about the direction (due to all ``yes'' answers for id questions), 
it chooses one of the neighbors as the next respondent uniformly at random.


\end{definition}


\medskip
In {\bf NA}-querying, Fig.~\ref{fig:Noninter} illustrates a
  candidate set of nodes inside a square, id/dir querying is performed in a batch with $r=1$.
  This hop-based candidate set selection has also been considered in \cite{Khim14,Choi17}, revealing that 
  it is a good approximation for the optimal one. 
  In {\bf AD}-querying, Fig.~\ref{fig:Inter} shows an example scenario that
  starting from the initial node, a sequence of nodes answer the
  queries truthfully or untruthfully for $r=1$.

\section{Main Results}
\label{sec:noninteractive}

We now present our main results which state the necessary and sufficient budgets to achieve $1-\delta$ detection accuracy for both querying types defined in the class of querying schemes $\mathcal{NA}(r,K)$ and $\mathcal{AD}(r,K),$ respectively. 

For presentational convenience, we define a Bernoulli random variable $X$ that represents a querier's answer for an id question, such that $X$ is one with probability (w.p.) $p$ and $X$ is zero w.p. $1-p$. Similarly, we define a querier's random answer $Y$ for a dir question, such that $Y$ is one w.p. $q$ and $Y$ is $i$ w.p. $(1-q)/(d-1)$, for $i = 2, \ldots, d.$ To abuse the notation, we use $H(p)$ and $H(q)$ to refer to the entropies of $X$ and $Y$, respectively. Throughout this paper, we also use the standard notation $H(\cdot)$ to denote the entropy of a given random variable or vector. 

\subsection{NA-Querying: Necessary and Sufficient Budgets}

\smallskip
\noindent{\bf \em (1) Necessary budget.}
We present an information theoretic lower bound
of the budget for the target detection probability $1-\delta$ inside the class of $\mathcal{NA}(r,K).$ We let $\mathcal{T}(r)=[T_1,T_2,\ldots, T_{\left\lfloor K/r \right\rfloor}]$ be the random vector where each $T_i$ is the random variable of infection time of the $i$-th node in the candidate set. Then, by appropriately choosing $r$, we have the following theorem.


\begin{theorem}
\label{theorem:lower}
Under $d$-regular tree $G$, as $N \rightarrow \infty,$ for any $0< \delta <1,$ there exists a constant $C=C(d),$ such that if
\begin{align}
\label{eqn:lower}
     K & \leq \frac{C \cdot H(\mathcal{T}(r^\star)) (2/\delta)^{1/2}}{f_{LN} (p,q)\log(\log (2/\delta))},
    \end{align}
where 
\begin{align}
f_{LN} (p,q) & =(1-H(p))+p(1-p)(\log_2 d -H(q)), \cr
r^\star & =  \left \lfloor 1+\frac{4 (1-p)\{7H(p)+2H(q)\}\log K}{3e\log(d-1)}\right\rfloor,
\end{align}
    then no algorithm in the class $\mathcal{NA}(r,K)$ can achieve the detection
probability $1-\delta.$ 
\end{theorem}

\smallskip
Note that $H(\mathcal{T}(r))$ can be expressed as a function of the diffusion rate $\lambda,$ see \cite{Sujay2012}. The implications of Theorem~\ref{theorem:lower} are in order. 
First, if the entropy $H(\mathcal{T}(r^\star))$ of the infection time is large,  the necessary amount of budget increases due to large uncertainty in figuring out a predecessor in the diffusion snapshot.  
Second, larger entropy for the answers of id/dir questions requires more
budget to achieve the target detection accuracy. Also, 
when $p$ goes to $1/2$ and $q$ goes to $1/d$, \ie, no information from the querying,
results in diverging the required budget (because $f_{LN}$ goes to zero).
Finally, if respondents are truthful in answering for the id question (\ie, $p=1$),
the direction answers does not effect the amount of necessary budget.

\begin{algorithm}[t!]
 \caption{{\bf MVNA}$(r)$}
\label{alg:noninteractive}
{\small

 \KwIn{Diffusion snapshot $G_N$, budget $K$, degree $d$, truthfulness probabilities
   $p>1/2$, $q>1/d.$} 
 \KwOut{Estimator $\hat{v}$}

\smallskip

$C_r  = S_I = S_D =\emptyset$\;
Choose the candidate set $C_r$ as in Definition~\ref{def:na} and ask the id/dir questions $r$ times to each node in $C_r$\;

\For{each $v \in C_r$}{
\textbf{Step1}: Count the number of `yes'es for the
identity question, stored at $\mu(v),$ and if $\mu(v)/r \geq 1/2$ then
add $v$ to $S_I$\;
\smallskip
\textbf{Step2}: For each of $v$'s neighbors, count the number of 
       designations for the dir question, choose the $v$'s neighbor, say $w$, with the largest count (under the rule of random tie breaking) as $v$'s `predecessor', and save a directed edge, called predecessor edge, $w \rightarrow v$
       \;
       \smallskip
}

Make a graph $G_{\text{pre}}$with all the predecessor edges and for each $v \in C_r,$ set $E(v) \leftarrow$ the number of all the descendants of $v$ 




$S_D \leftarrow \arg\max_{v\in C_r}
   |E(v)|$\;
 \If{$S_I \cap S_D=\emptyset$}{
 If $p=1$, set $ \hat{v} \leftarrow \arg\max_{v\in S_I }
   \mathbb{P}(G_N | v=v_1)$ otherwise, set
 $ \hat{v} \leftarrow \arg\max_{v\in S_I \cup S_D}
   \mathbb{P}(G_N | v=v_1)$\;
   } \Else{ $ \hat{v} \leftarrow \arg\max_{v\in S_I \cap S_D}
   \mathbb{P}(G_N | v=v_1)$\; }
 Return $\hat{v}$\;
}
\end{algorithm}

\smallskip
\noindent{\bf \em (2) Sufficient budget.}
To compute a sufficient budget, a natural choice would be to use the MLE (Maximum Likelihood Estimator), which, however, turns out to be computationally intractable for large $N$ due to too much randomness of the diffusion snapshot and query answers. Hence, we consider a simple estimation 
algorithm named {\bf MVNA}$(r)$ that is based on 
majority voting for both the id and dir questions. To briefly explain how the algorithm behaves, we first select the candidate set $C_r$ of size $\left\lfloor K/r \right\rfloor$ that has the least hop-distance from the RC, then we ask $r$ times of id/dir questions to each node in the candidate set (Line 1). Then, we filter out the nodes that are more likely to be the source and save them in $S_I$ (Line 4) and using the results of the dir questions, compute $E(v)$ that correspond to how many nodes in $C_r$ hints that $v$ is likely to be the source node (Lines 5 and 6). 
Finally, we
choose a node with maximal likelihood in $S_I \cap S_D$ and if
$S_I \cap S_D = \emptyset,$ we simply perform the same task for $S_I \cup S_D.$
It is easy to see that the time complexity is $O(\max\{N,K^2 /r\})$.

Now, Theorem~\ref{theorem:noninteractive} quantifies the amount of querying
budget that is sufficient to obtain arbitrary detection probability by
appropriately choosing the number of questions to be asked. 

%
%
%


\smallskip
\begin{theorem}
\label{theorem:noninteractive}
For any $0< \delta <1,$ the detection
probability under $d$-regular
tree $G$ is at least $1-\delta,$ as $N \rightarrow \infty,$ if
\begin{align}
\label{eqn:elower}
     K & \geq \frac{12d/(d-2)(2/\delta)}{f_{N} (p,q)\log(\log (2/\delta))},
    \end{align}
    where $f_{N} (p,q)=3(p-1/2)^2 + \frac{(d-1)p(1-p)}{3d} (q-1/d)^2 $ under {\bf MVNA}$(r^\star),$ where
    \begin{align*}
r^\star &= \left \lfloor 1+\frac{2(1-p)\{1+(1-q)^2\}\log K}{e\log(d-1)}\right\rfloor.
\end{align*}
\end{theorem}

\smallskip

We briefly discuss the implications of the above theorem. 
First, we see that $(1/\delta)^{1/2}$ times more budget is required that the necessary one, which is because we consider a simple, approximate estimation algorithm. 
Second, the dir question does not effect the
sufficient budget $K$ if $p=1$ \ie, no untruthfulness
for the id question as in Theorem~\ref{theorem:lower}. However, if $p<1$,
the information from the answers for the dir questions reduces the sufficient
amount of budget, because $f_N$ increases in the denominator of \eqref{eqn:elower}.
Finally, when $p$ goes to $1/2$ and $q$ goes to $1/d$, the required budget
diverges due to the lack of information from the querying.




\vspace{-0.18cm}
\subsection{AD-Querying: Necessary and Sufficient Budgets}
\label{sec:interactive}
\vspace{-0.08cm}
\noindent{\bf \em (1) Necessary budget.}
Next, we present an information theoretic lower bound
of the budget for the target detection probability $1-\delta$ for the algorithms in the class $\mathcal{AD}(r,K)$ in Theorem~\ref{theorem:lowerad} by choosing $r$, appropriately.

\smallskip
\begin{theorem}
\label{theorem:lowerad}
Under $d$-regular tree $G$, as $N \rightarrow \infty,$ for any $0< \delta <1,$ there exists a constant $C=C(d),$ such that if
\begin{align}
\label{eqn:lowerad}
     K & \leq \frac{C \cdot H(\mathcal{T}(r^\star)) (\log(7/\delta))^{\alpha/2}}{f_{LA} (p,q)\log(\log (7/\delta))},
    \end{align}
    for $\alpha=2$ if $p <1$ and $\alpha=1$ if $p=1$ where
\begin{align}
&f_{LA} (p,q) =(1-H(p))+p(\log_2 d -H(q)), \cr
& r^\star  =  \left \lfloor 1+\frac{7d  p\{3H(p)+2dH(q)\}\log \log K}{2(d-1)}\right\rfloor,
\end{align}
   then no algorithm in the class $\mathcal{AD}(r,K)$ can achieve the detection
probability $1-\delta.$ 
\end{theorem}

\smallskip
We describe the implications of Theorem~\ref{theorem:lowerad} as follows.
First, when $p$ goes to $1/2$ and $q$ goes to $1/d$, \ie, no information from the querying causes diverging the required budget (because $f_{LA}$ becomes zero). Second, the positive untruthfulness for the id question ($p<1$) requires $\log^{1/2} (1/\delta)$ times more budget than that under the perfect truthfulness ($p=1$). This is because more sampling is necessary to learn the source from the answers of the id questions when $p <1$, whereas no such learning is required for finding the source when $p=1$. Third, large truthfulness (\ie, large $p$) gives more chances to get the direction answers which decreases the amount of budget. Finally, we see that the order is reduced from $1/\delta$ to $\log(1/\delta)$, compared to that in Theorem~\ref{theorem:lower}.

\begin{algorithm}[t!]
{\small
 \KwIn{Diffusion snapshot $G_N$, querying budget $K$, degree $d$, truthful
   probabilities $p>1/2$, $q>1/d$} 
   \KwOut{Estimated rumor source $\hat{v}$}

\smallskip
$S_I = S_D = \emptyset$ and $\eta (v)= 0$ for all $v \in V_N$\;
Set the initial node $s$ by RC\;
\While{$K\geq r$}{
    {\If{$p=1$}{ If $s=v_1$, return $\hat{v}=s$ otherwise, go to step 2\;}\Else{
    \textbf{Step1}: Set $\eta (s) \leftarrow \eta (s)+1$ which describes that the node $s$ is taken as a respondent and count the number of ``yes''es for the
identity question, stored at $\mu(s)$, and if $\mu(v)/r \geq 1/2$ then
add $v$ to $S_I$\; 
}
\smallskip
   \textbf{Step2}: Count the number of
       ``designations'' for the
       direction question among $s$'s neighbors, and choose the largest
       counted node as the predecessor with a random tie breaking\;
       Set such chosen node by $s$ and $K \leftarrow K-r$\;
      }  { } }
$S_D \leftarrow \arg\max_{v\in V_N}
   \eta (v)$\;
    \If{$S_I \cap S_D=\emptyset$}{ $ \hat{v} \leftarrow \arg\max_{v\in S_I \cup S_D}
   \mathbb{P}(G_N | v=v_1)$\; } \Else{ $ \hat{v} \leftarrow \arg\max_{v\in S_I \cap S_D}
   \mathbb{P}(G_N | v=v_1)$\; }
 Return $\hat{v}=s$\;
}
\caption{{\bf MVAD}$(r)$}
\label{alg:interactive}
\end{algorithm}


\noindent{\bf \em (2) Sufficient budget.}
In {\bf AD}-querying, due to the similar computational issue to {\bf NA}-querying in using the MLE, we also consider a simple estimation algorithm 
to obtain a sufficient budget named by {\bf MVAD}$(r),$ which is again based on majority voting for both the id and dir questions.
In this algorithm, we choose the RC as the initial node and perform different querying procedures for the following two cases: (i) $p=1$ and (ii) $p<1$. First, when $p=1$, since there is no untruthfulness of the answers of the id questions, we check whether the current respondent $s$ is the source or not. If yes, then the algorithm is terminated and it outputs the node $s$ as a result (Line 5). If not, it asks of $s$ the dir question $r$ times and chooses one predecessor by majority voting with random tie breaking (Line 8). Then, for the chosen respondent, we perform the same procedure until we meet the source or the budget is exhausted. Second, when $p <1,$ we first add one in $\eta (s)$ which is the count that the node $s$ is taken as the respondent. Next, due to untruthfulness, we count the number of ``yes'' answers for the id question and apply majority voting to filter out the nodes that are highly likely to be the source and save them in $S_I$ (Line 7). For the negative answers for id questions, we count the designations of neighbors and apply majority voting to choose the next respondent. Then, we perform the same procedure to the chosen node and repeat this until the budget is exhausted. To filter out more probable source node from the direction answers, we compare the number that is taken as the respondent by designation from the neighbors in $\eta (v),$ and we choose the node which has the maximal count of it and save them into $S_D$ (Line 10). Finally, we
select a node with maximal likelihood in $S_I \cap S_D$ or $S_I \cup S_D$ (Lines 11-14). We easily see that the time complexity of this algorithm is $O(\max\{N,K\})$.
Now, Theorem~\ref{theorem:interactive} quantifies the sufficient amount of budget
to obtain arbitrary detection probability by
appropriately choosing the number of questions to be asked. 
\smallskip
\begin{theorem}
\label{theorem:interactive}
For any $0< \delta <1,$ the detection
probability under $d$-regular
tree $G$ is at least $1-\delta,$ as $N \rightarrow \infty,$ if
\begin{align}
\label{eqn:elower1}
    K\geq \frac{2(2d-3)/d(\log(7/\delta))^{\alpha}}{f_{A} (p,q)\log (\log (7/\delta))},
    \end{align}
   where $f_{A} (p,q)=\frac{2d}{d-1} (p-1/2)^2 + \frac{d-1}{d-2} (q-1/d)^3$ and
     $\alpha=2$ if $p <1$ and $\alpha=1$ if $p=1$ under {\bf MVAD}$(r^\star)$, where 
       \begin{align*}
r^\star &= \left\lfloor1+\frac{7d^2 \{2(1-p)^3 + (1-q)^2\} \log\log
  K}{3(d-1)}\right\rfloor.
\end{align*}
\end{theorem}

\smallskip

The gap between necessary and sufficient budgets is $\log(1/\delta)$ when $p<1,$ and $\log^{1/2} (1/\delta),$ when $p=1.$ 
Note that we have $\log(1/\delta)$ factor reduction from what is sufficient under {\bf MVNA}$(r^\star)$ in the non-adaptive case. 
Further, as expected, we see that the sufficient
budget arbitrarily grows as $p$ goes to $1/2$ and $q$ goes to $1/d$, respectively.

\vspace{-0.1cm}
\subsection{Adaptivity Gap: Lower and Upper Bounds}
\label{sec:adaptive}
Using our analytical results stated in Theorems~\ref{theorem:lower}-\ref{theorem:interactive}, we now establish the quantified adaptivity gap defined as follows: 

\smallskip
\begin{definition}(Adaptivity Gap)
Let $K_{na}(\delta)$ and $K_{ad}(\delta)$ be the amount of budget needed to obtain $(1-\delta)$ detection probability for $0< \delta <1$ by the optimal algorithms in the classes $\mathcal{NA}(r,K)$ and $\mathcal{AD}(r,K),$ respectively. Then, the adaptivity gap, $\text{AG}(\delta)$ is defined as $K_{na}(\delta)/K_{ad}(\delta).$
\end{definition}
\smallskip


\smallskip
\begin{theorem}
\label{thm:AG}
For a given $0<\delta<1,$ there exist a constant $r$ and two other constants $U_1 = U_1(r,p,q)$ and $U_2 = U_2(r,p,q)$, where 
the constant $r$ corresponds to the number of repeated id/dir questions for each respondent in both classes $\mathcal{NA}(r,K)$ and $\mathcal{AD}(r,K)$, such that 
\begin{align}
\label{eqn:elower2}
   \frac{U_1 \cdot (1/\delta)^{1/2}}{\log^{\alpha} (1/\delta)} \leq AG(\delta) \leq \frac{U_2 \cdot (1/\delta)}{\log^{\alpha/2} (1/\delta)},
    \end{align}
   where $\alpha=2$ if $p <1$, and $\alpha=1$ if $p=1$.

\end{theorem}

\smallskip
In Theorem~\ref{thm:AG}, we see that
for a given target detection probability $1-\delta$, 
the required amount of querying budget by adaptive querying
asymptotically decreases from $(1/\delta)$ to $\log (1/\delta),$
This implies that there is a significant gain of querying 
in the adaptive manner. 
Further, the difference of upper and lower bounds of $\text{AG}(\delta)$ is expressed by square root in our algorithm classes, when we use {\bf MVNA}$(r^\star)$ and {\bf MVAD}$(r^\star)$ for sufficient budgets, respectively.

\section{Proofs}\label{sec:proof}
In this section, we will provide the
proofs for the Theorems. The whole proof will be provided in our supplementary material \cite{Jae16}.

\subsection{Proof of Theorem~\ref{theorem:lower}}
For a given $r,$ we introduce the notation $V_l,$ which is
equivalent to $C_r,$ where the hop distance $l=\frac{\log\left(\frac{K(d-2)}{rd}+2\right)}{\log(d-1)}$.
 Also for notational simplicity, we simply use $
\prob{\hat{v}=v_{1}}$ to refer to $\lim_{N \to \infty }\prob{\hat{v}(G_N,r)=v_{1}}$
for any estimator given the snapshot $G_N$ and redundancy parameter $r$ in the proof section.
Then, the detection probability is expressed as the product of the two
terms:
    \begin{align}
      \label{eqn:detect0}
      \prob{\hat{v} = v_{1}} &=   \prob{v_1 \in  V_{l}}\times  \prob{\hat{v}=v_1|v_1 \in V_{l}},
    \end{align}
where the first one is the probability that the source is in the $l$-hop based candidate set $V_l$
and the second term is the probability that the estimated node is exactly the source in the candidate set for
any learning algorithm under the algorithm class $\mathcal{C}(l,r)$.
We first obtain the upper bound of probability of first term in \eqref{eqn:detect0}
in the following lemma.

\smallskip
\begin{lemma}\label{lem:multihop_upper}
For $d$-regular trees,
\begin{equation}
\prob{v_1 \in V_{l}} \leq 1-c\cdot e^{-l\log l},
\label{eqn:multihop_upper}
\end{equation}
where $c = 4d/3(d-2)$.
\end{lemma}
\smallskip

We will closely look at the case of each $l$, to derive the
probability that the rumor center $v_{RC}$ is exactly $l$-hop distant
from the rumor source $v_1$. Let $\delta_1$ be the error for the $\prob{v_1 \notin V_{l}}$ then
it is lower bounded by $\delta_1 \geq c\cdot e^{-l\log l}$.

To obtain the second term in \eqref{eqn:detect0}, we use the information theoretical techniques
for the direct graph inference as done in \cite{Sujay2012} with partial observation because, if the rumor spread from the source
we can obtain a direct tree where all direction of edges are outgoing from the source.
From the assumption
of independent answers of queries, we see that the snapshot from one querying process with untruthful for direction question is equivalent to the
snapshot of diffusion flow from the source under the IC-diffusion model with noisy observation.
By using these fact and the result of graph learning techniques from the epidemic cascades in \cite{Sujay2012}, we obtain the following lemma.

\smallskip
\begin{lemma}\label{lem:information}
For any graph estimator to have a probability of error of $\delta_2>0$, it needs $r$ queries to the candidate
set $V_l$ with $|V_l|=n$ that satisfies
\begin{equation}\label{eqn:information1}
r\geq \frac{\log(1/\delta_2)H(T) (n-1)\log \frac{n}{2}}{ n((1-H(p))+p (1-p)(\log_2 d-H(q)))},
\end{equation}
where $H(T)$ is the entropy of infection time vector and $H(p)=p \log p + (1-p)\log (1-p)$ and $H(q)= q \log q + (1-q)\log \frac{1-q}{d-1} $, respectively.
\end{lemma}
\smallskip

This result indicates that if there is no information from query, \ie, $p=1/2$ and $q=1/d$, the required number of queries diverges. Further, if the uncertainty of infection time $H(T)$ for the nodes in $V_l$ increases, the required queries also increases. Then, from the disjoint of two error event and by setting $\delta_1 = \delta_2 =\delta /2$ with $l=\log\left(\frac{K(d-2)}{rd}+2\right)/\log(d-1)$, we have
  \begin{align*}
     \prob{&\hat{v} \neq v_{1}} \geq c\cdot e^{-\frac{\log\left(\frac{K}{r}\right)}{\log(d-1)}\log \frac{\log\left(\frac{K}{r}\right)}{\log(d-1)}} \cr
     & +
     e^{-\frac{H(T)(\frac{K}{r}-1)\log \frac{K}{2r}}{K((1-H(p))+p (1-p)(\log_2 d-H(q)))}}
     \geq \delta.
    \end{align*}
From the fact that $\lambda =1$ in our setting and Lemma 2 in \cite{Sujay2012}, we obtain $H(T)\leq K/r $ and by differentiation of above lower bound with respect to $r$, we approximately obtain
$r^\star= \left \lfloor 1+\frac{4 (1-p)\{7H(p)+H(q)\}\log K}{3e\log(d-1)}\right\rfloor$ where
the derivation is given in the supplementary material.
Since if we use the $r^\star$, it gives the upper bound of detection probability hence,
we put it to the obtained upper-bound which is expressed as a
function of $K,$ as follows:
\begin{align}
  \label{eq:kkk1}
  &\prob{\hat{v} \neq v_{1}}  \cr
  & \geq  \frac{1}{2}\ e^{-h_1 (T,p,q) \log K\log (\log K)}+\frac{c}{4} e^{- 2 h_1 (T,p,q)\log K\log (\log K)}\cr
&   \geq  C_d e^{-2 h_1 (T,p,q)\log K \log (\log K)},
\end{align}
where $C_d=(c+3)/4$ and $h_1 (T,p,q)= H(T)^{-1}(1-H(p))+p(1-p)(\log_2 d-H(q))$. If we set $\delta \leq C_d e^{-2 h_1 (T,p,q)\log K\log (\log K)},$ we find the value $K$ such that its assignment to
\eqref{eq:kkk1} produces the error probability $\delta,$ and we finally obtain the
desired lower-bound of $K$ as in Theorem~\ref{theorem:lower}.

\subsection{Proof of Theorem~\ref{theorem:noninteractive}}

We first provide the lower bound on detection probability of {\bf
  MVNA}$(r)$ for a given $K$ and $r$ in the following lemma.

\smallskip
\begin{lemma}
  \label{lem:simple_prob}
  For $d$-regular trees ($d\geq 3$), a given budget
  $K,$ our estimator $\hat{v}$ from {\bf MVNA}$(r)$ has the
  following lower-bound of the detection probability:
\begin{multline}
  \label{eqn:detect}
\prob{\hat{v}=v_{1}}\geq 1-c\left(\frac{r+p+q}{r+2}\right)^3\cdot
  \exp\Bigg( \frac{-h_d
  (K,r)w_d (p,q)}{2}\Bigg),
\end{multline}
where $c = 7(d+1)/d$ and $w_d (p,q)=\frac{1}{2}(4(p-1/2)^2+(d/(d-1))^3 (q-1/d)^3)$.
The term $h_d (K,r)$ is given by
$$h_d (K,r):=\frac{\log\left(\frac{K}{r}\right)}{\log(d-1)}\log\left(
\frac{\log\left(\frac{K}{r}\right)}{\log(d-1)}\right).$$
\end{lemma}
\smallskip

\begin{proof}
Under the {\bf MVNA}$(r)$, the detection probability is expressed as the product of the three
terms:
    \begin{align}
      \label{eqn:detection}
      \prob{\hat{v} = v_{1}} &=   \prob{v_1 \in  V_{l}}\times  \prob{\hat{v}=v_1|v_1 \in V_{l}} \cr
                               & = \prob{v_1 \in  V_{l}} \times \prob{v_1
                                 \in \hat{V}|v_1 \in V_{l}}   \cr
                               &  \times \prob{v_1 =v_{LRC}|v_1 \in \hat{V}},
    \end{align}
where $\hat{V}:=S_I \cap S_D$ if it is not empty or $\hat{V}:=S_I \cup S_D$, otherwise. This is the filtered candidate set in {\bf
  MVNA}$(r)$ and  $v_{LRC}$ is the node in $\hat{V}$ that has the
highest rumor centrality \ie, likelihood, where $LRC$ means the local rumor center.
We will drive the lower bounds of the first, second, and the third terms
of RHS of \eqref{eqn:detection}.
The first term of RHS of \eqref{eqn:detection} is bounded by
\begin{align}
  \label{eq:first}
  \prob{v_1 \in V_{l}} \geq 1-c\cdot e^{-(l/2)\log l},
\end{align}
where the constant $c = 7(d+1)/d$ from Corollary 2 of \cite{Khim14}.
Let $S_N$ be the set of revealed nodes itself as the rumor source and let $S_I$ be the set of nodes which minimizing the errors.
If the true source is in $V_{l},$ then the probability that it is most indicated node for
a given budget $K$ with the repetition count $r$ and truth probability $p>1/2$ and $q>1/d$ is given by
    \begin{align}\label{eqn:d}
      \prob{v_1 &=v_{LRC}|v_1 \in \hat{V}}\cr
     & =\prob{v_1= \arg\max_{v\in S_I\cap S_D}R(v,G_{N})|K,p,q}.
    \end{align}
To obtain this, we consider that
if $p>1/2$,
 the probability $v_1 \in S_I$ by the majority voting, because the selected node can be designation again
in the algorithm. We let total number of queries by $r\geq1$, we let $W=\sum_{i=1}^{r}X_i (v_1)$ for the source node $v_1$, then
the probability that true source is in the filtration set $S_I$ is given by $\prob{W\geq r/2}=\sum_{j=0}^{\lfloor r/2\rfloor}\binom{ r}{j}(1-p)^{j}p^{r-j}.$
Then, from this relation, we have the following lemmas
whose proofs are will be provided in \cite{Choi17}:
\smallskip
\begin{lemma}(\cite{Choi17})
When $p>1/2,$
\begin{eqnarray*}
\prob{v_1 \in S_I|v_1 \in V_{l}}  & \geq &p+(1-p)(1-e^{-(p-1/2)^{2}\log r}).
\end{eqnarray*}
\label{lem:majority}
\end{lemma}
This result implies the lower bound of probability that the
source is in $S_I$ for a given $r$. Next, we will obtain the probability that the source is in $S_D$ after filtration of the direction answers.
To do this, we first consider that the total number of direction queries $N_d$ is a random variable which is given by:
\begin{align*}
P(N_d = k)=\begin{cases}
\binom{r}{k}p^{r -k}(1-p)^{k}&\mbox{if}~v=v_1\\
\binom{r}{k}(1-p)^{r -k}p^{k}&\mbox{if}~v \neq v_1,
\end{cases}
\end{align*}
where $k$ is less than parameter $r$. Using this fact, we obtain the following result.
\smallskip

\begin{lemma}
When $p>1/2$ and $q>1/d$,
\begin{eqnarray*}
\prob{v_1 \in S_D|v_1 \in V_{l}}  & \geq & 1- e^{-\frac{rp(d-1)(q-1/d)^2}{3d}}.
\end{eqnarray*}
\label{lem:consistence}
\end{lemma}
\smallskip

This result shows the lower bound of probability that the source is in $S_I$ for a given $r$.
By considering the two results in the above, we have the following lemma.

\smallskip
\begin{lemma}
For given repetition count $r$, we have
\begin{equation}\label{eqn:d1}
    \begin{aligned}
      P(v_1 \in S_I \cap S_D|v_1 \in V_l) \geq 1-2e^{-f(p,q) 2r\log r}
    \end{aligned}
  \end{equation}
where $f (p,q)=3(p-1/2)^2 + \frac{d-1}{3d} p(1-p)(q-1/d)^2$.
\label{lem:finalfil}
\end{lemma}
\smallskip

Then, we obtain the following lemma, which is the lower bound of detection probability
among the final candidate set.

\smallskip
\begin{lemma} \label{lem:filtration}
  When $d\geq3$, $p>1/2$ and $q>1/d$,
  \begin{eqnarray*}
    \prob{v_1 =v_{LRC}|v_1 \in S_I \cap S_D}  & \geq & 1-e^{-f(p,q) r\log r}.
  \end{eqnarray*}
\end{lemma}
\smallskip

Merging these lower-bound with the lower-bound in
\eqref{eq:first} where we plug in
$l=\frac{\log\left(\frac{K(d-2)}{rd}+2\right)}{\log(d-1)},$ we finally get the lower bound of detection probability
 of {\bf MVNA}$(r)$ for a given repetition count $r$ and this completes the proof of
 Lemma~\ref{lem:simple_prob}.
\end{proof}

To finish the proof of theorem, note that the second term of RHS of \eqref{eqn:detect} is the probability that the
source is in the candidate set for given $K$ and $r$. Hence, one can see
that for a fixed $K$, large $r$ leads to the decreasing detection
probability due to the smaller candidate set.  However, increasing $r$
positively affects the first term of RHS of \eqref{eqn:detect}, so that
there is a trade off in selecting a proper $r$.
By derivation of the result with respect to $r$, we first obtain $r^\star$ which maximizes the detection
probability by $r^\star= \left \lfloor 1+\frac{2(1-p)\{1+(1-q)^2\}\log K}{e\log(d-1)}\right\rfloor$ in {\bf MVNA}$(r^\star)$
and put this into the error probability $\prob{\hat{v}\neq v_{1}}$ such as

\begin{align}
  \label{eq:ppp}
  \prob{\hat{v}&\neq v_{1}} \leq e^{-f(p,q) r\log r}+2e^{-f(p,q) 2r\log r}+c\cdot e^{- \frac{l}{2}\log l},
\end{align}
where the constant $c$ is the same as that in \eqref{eq:first}.
Now, we first put
$l=\frac{\log\left(\frac{K(d-2)}{rd}+2\right)}{\log(d-1)}$ into
\eqref{eq:ppp} and obtained the upper-bound of \eqref{eq:ppp}, expressed
as a function of $r,$ for a given $p$ and $q$ and the constant $c.$
Then, we take $r^*$ and put it to the obtained upper-bound which is expressed as a
function of $K,$ as follows:
\begin{align}
  \label{eq:kkk2}
  \prob{\hat{v} \neq v_{1}}  & \leq  3\ e^{-f(p,q) \log K\log (\log K)}+c e^{- \frac{\log K}{2}\log (\log K)}\cr
&   \leq  c_1 e^{-f (p,q)\frac{\log K}{2} \log (\log K)},
\end{align}
where $c_1=c+3$. If we set $\delta \geq c_1 e^{-f (p,q)\frac{\log K}{2}\log (\log K)},$ we find the value of $K$ such that its assignment to
\eqref{eq:kkk2} produces the error probability $\delta,$ and we get the
desired lower-bound of $K$ as in the theorem statement. This completes
the proof of Theorem~\ref{theorem:noninteractive}.

\subsection{Proof of Theorem~\ref{theorem:lowerad}}
We will show the lower bound for given $K$ and $r$ of the case $p<1$. \footnote{The result
  for $p=1$ is similar to this except the termination of querying process when it meets the source.}
For a given $r,$ we let $V_L$ be the set of all infected nodes from the rumor center within a distance $L:=K/r$ then we see that
the querying dynamic still becomes a directed tree construction rooted by the source $v_1$.
Then, the detection probability is expressed as the product of the two
terms:
    \begin{align}
      \label{eqn:detect01}
      \prob{\hat{v} = v_{1}} &=   \prob{v_1 \in  V_{L}}\times  \prob{\hat{v}=v_1|v_1 \in V_{L}},
    \end{align}
where the first one is the probability that the distance between source and rumor center is less than $K/r$
and the second term is the probability that the estimated node is exactly the source in the candidate set for
any learning algorithm under the algorithm class $\mathcal{AD}(r,K)$.
First, from Lemma~\ref{lem:multihop_upper}, we have that the probability of first term in \eqref{eqn:detect01} is
upper bounded by $1-ce^{-(K/r)\log(K/r)}$ where $c=4d/3(d-2)$ for a given budget $K$ and repetition count $r$.
We see that
the querying dynamic still becomes a directed tree construction rooted by the source $v_1$.
However, different to the NA-querying, the querying process gives direction data of a subgraph of the original direct tree
because the querier chooses a node, interactively.
For a given $r$, let $Z_{r,i}$ be the answer data of querying for a selected queried node $i$ where $1 \leq i \leq K/r.$ Then,
from the assumption of the algorithm class $\mathcal{AD}(r,K)$,
the joint entropy for the random answers with the infection time random vector $T$,
$H(T, Z_{r,1},\ldots,Z_{r,K/r})$ is given by
 \begin{align}\label{eqn:entropy}
     H(T, Z_{r,1},&\ldots,Z_{r,K/r})=\sum_{i=1}^{K/r}H(T, Z_{r,i}|Z_{r,i-1},\ldots,Z_{r,1})\cr
     &=\sum_{i=1}^{K/r}H(T, Z_{r,i}|Z_{r,i-1})\stackrel{(a)}{=}\sum_{i=1}^{K/r}H(T, Z_{r,i}),
    \end{align}
where $(a)$ is from the fact that all data $Z_{r,i}$ are independent.
Let $G^*$ be the true directed graph and let $\hat{G}$ be be an estimated directed tree from the
sequential answers of adaptive querying $(Z_{r,1},\ldots,Z_{r,K/r})$. Then, we see that
this defines a Markov chain
$$G^* \rightarrow (T, Z_{r,1},\ldots,Z_{r,K/r}) \rightarrow \hat{G},$$
from the defined algorithm class $\mathcal{AD}(r,K)$. By property of the mutual information, we have
 \begin{align}\label{eqn:entropy1}
    I(G^*; &T, Z_{r,1},\ldots,Z_{r,K/r})\cr
    &\leq H(T, Z_{r,1},\ldots,Z_{r,K/r})=\sum_{i=1}^{K/r}H(T, Z_{r,i}) \cr
    &\stackrel{(a)}{=} (K/r)H(T, Z_{r,1})\cr
    &\stackrel{(b)}{\leq} (KH(T)/r)[r(1-H(p))+rp(\log_2 d - H(q))]\cr
    &= KH(T)[(1-H(p))+p(\log_2 d - H(q))]\cr
    &:=Kh(p,q),
    \end{align}
where $(a)$ follows from the fact that the answers $Z_{r,i}$ are mutually exclusive
and $(b)$ is from the fact that $H(T, Z_{r,1})=(1-H(p)+rp(\log_2 d - H(q)))/H(T)$ since
the number of direction answers follows binomial distribution. Let $\mathcal{G}_{K/r}$ be the
set of possible directed tree in $V_s$ then we have $|\mathcal{G}_{K/r}|\leq (K/r)\log (K/2r)$.
Using the Fano's inequality on the Markov chain $G^* \rightarrow (Z_{r,1},\ldots,Z_{r,K/r}) \rightarrow \hat{G},$
we obtain
 \begin{align}\label{eqn:entropy2}
   \prob{G \neq G^*}
   &\geq  \frac{I(G^*; Z_{r,1},\ldots,Z_{r,K/r})+h(p,q)}{H(T)\log|\mathcal{G}_{K/r}|}\cr
   & \geq  \frac{Kh(p,q)+h(p,q)}{\frac{KH(T)}{r}\log (\frac{K}{2r}-1)}.
    \end{align}
From the disjoint of two error event and by setting $\delta_1 = \delta_2 =\delta /2$ for each error, we have
  \begin{align}\label{eqn:entropy3}
     \prob{&\hat{v} \neq v_{1}} \geq c\cdot e^{-(K/r)\log(K/r)} \cr
     & +e^{-\frac{\frac{KH(T)}{r}\log (\frac{K}{2r}-1)}{Kh(p,q)+h(p,q)}}
     \geq \delta.
    \end{align}
From the fact that $\lambda =1$ in our setting and Lemma 2 in \cite{Sujay2012}, we approximately obtain $H(T)\leq K/r $ and
by differentiation of above lower bound with respect to $r$, we obtain
$r^\star= \left \lfloor 1+\frac{7 dp\{3H(p)+2dH(q)\}\log \log K}{2(d-1)}\right\rfloor$ where
the derivation is given in the supplementary material.
Since if we use the $r^\star$, it gives the upper bound of detection probability hence,
we put it to the obtained upper-bound which is expressed as a
function of $K,$ as follows:
\begin{align}
  \label{eq:kkk3}
  \prob{\hat{v} \neq v_{1}}  & \geq  \frac{1}{3}\ e^{-h_2(T,p,q) K\log (\log K)}+\frac{c}{4} e^{- 7 h_2 (T,p,q)K\log (\log K)}\cr
&   \geq  C_d e^{-7h_2 (T,p,q) K\log (\log K)},
\end{align}
where $C_d=2(c+3)/7$ and $h_2 (T,p,q)= H(T)^{-1}(1-H(p))+(1-p)(\log_2 d-H(q))$. If we set $\delta \leq C_d e^{-7 h_2 (T,p,q)K\log (\log K)},$ we find the value of $K$ such that its assignment to
\eqref{eq:kkk3} produces the error probability $\delta,$ and we get the
desired lower-bound of $K$ as in the theorem statement.
Then, we finally obtain the result and this completes the proof of
Theorem~\ref{theorem:lowerad}.

\subsection{Proof of Theorem~\ref{theorem:interactive}}
We will show the lower bound on the detection
  probability for given $K$ and $r$ of the case $p<1$ \footnote{The result
  for $p=1$ is given in \cite{Choi17} and we omit it here.} in Lemma~\ref{lem:inter_prob}.

\smallskip
\begin{lemma}
  \label{lem:inter_prob}
  For $d$-regular trees ($d\geq 3$), a given budget
  $K,$ our estimator $\hat{v}$ from {\bf MVAD}$(r)$ has the
  detection probability lower-bounded by:
\begin{align}
  \label{eqn:detect2}
\prob{\hat{v}=v_{1}} \ge & 1-c (g_d (r,q))^3 \cr
&\cdot \exp\left [-\left(p-\frac{1}{2}\right)^2\left(\frac{K}{r}\right)\log \left(\frac{K}{r}\right) \right],
\end{align}
where $g_d (r,q):=  e^{-\frac{r(d-1)(q-1/d)^2}{3d(1-q)}}$ and
$c=(5d+1)/d$.
\end{lemma}
\smallskip

\begin{proof}
For the {\bf MVAD}$(r)$, for a given $r,$ we introduce the notation $V_s,$
where the set of all queried nodes of the algorithm. From the initial queried node, we need the probability
that the source is in the set of queried node by some policy $P \in \mathcal{P}(v_I)$.
Then, the detection probability is also expressed by the product of the three
terms:
    \begin{align}
      \label{eqn:detection1}
      \prob{\hat{v} = v_{1}} &=   \prob{v_1 \in  V_{L}}\times  \prob{\hat{v}=v_1|v_1 \in V_{L}} \cr
                               & = \prob{v_1 \in  V_{L}} \times \prob{v_1
                                 \in \hat{V}|v_1 \in V_{L}}   \cr
                               &  \times \prob{v_1 =v_{LRC}|v_1 \in \hat{V}},
    \end{align}
where $V_{L} = \{v| d(v_{RC},v) \le K/r\}$ because the number of budget is $K$ and $\hat{V}=S_I \cap S_D$
if it is not empty or $\hat{V}=S_I \cup S_D$, otherwise.
From the result in Corollary 2 of \cite{Khim14}, we have $\prob{E_1}\leq c \cdot e^{- (K/r)\log K/r}$
since we use additional direction query with identity question. For the second part of probability in \eqref{eqn:detection1},
we obtain the following lemma.

\smallskip
\begin{lemma}
When $p>1/2,$
\begin{equation*}
 \begin{aligned}
\prob{v_1 &\in S_I|v_1 \in V_{L}}  \\
&\geq \left(p+(1-p)(1-e^{-(p-1/2)^{2}\log r})\right) \left(1-ce^{-\frac{Kp}{r}  (q-1/d)^3}\right).
   \end{aligned}
   \end{equation*}
\label{lem:idinter}
\end{lemma}
\smallskip

\begin{proof}
Let $Q_K (v)$ be the number of queries to a node $v \in V_l$ when there are $K$ queries then we have
\begin{equation*}
    \begin{aligned}
     &\prob{Q_K (v_1)\geq1}\\
     &=\sum_{i=1}^{l}\prob{Q_K (v_1)\geq1|d(v_1, v_{RC})=i}\prob{d(v_1, v_{RC})=i}.
    \end{aligned}
  \end{equation*}
where $\prob{d(v_1,v_{RC}) = i}$ is the probability that the distance from the rumor center to rumor source is $i$ and this probability become smaller if the distance between rumor source and rumor center is larger.
From this, we have the following result for the lower bound of the probability of distance between the rumor center and source.
\smallskip
\begin{proposition}
For $d$-regular trees,
\begin{equation}
\prob{d(v_1,v_{RC}) = i}\geq \left(\frac{d-1}{d}\right)^{i} e^{-(i+1)}.
\label{eqn:interactive_multihop1}
\end{equation}
\label{pro:distance1}
\end{proposition}
\smallskip

Next, we construct the following Markov chain.
Let $\hat{p}:=\prob{W = r}$ for the identity questions \ie, there is no ``no'' answers for the
identity questions so that the algorithm should chooses one of neighbor nodes uniformly at random
and let $\hat{q}:=\prob{Z_1 (v) > Z_j
(v),~ \forall j}$ for the direction question, respectively.
Different to the case for $p=1$ which the node reveals itself as the rumor source or not
with probability one so that the Markov chain has the absorbing state, in this case, there is no such a state.
To handle this issue, we use
the information that how many times the neighbors indicate a node as its parent and how many times a node reveals itself as the rumor source.
To do that, we consider the case that there is a \emph{token}\footnote{The token keeper is regarded as the current respondent in this model.} from the initial state and
it move to the next state after additional querying follows the answer. Then this probability is the same that after $K/r$-step of Markov chain, and we expect that the rumor source $v_1$ will have the largest chance to keeping this token due to the assumption of biased answer. Let $X_n$ be the state (node) which keep this token at time $n$ where
the state is consist of all node in $V_N$. The initial state is the rumor center such that $X_0 =0$ where $0$ indicates the rumor center. Then there are $(d(d-1)^{K/r}-2)/(d-2)$ states and we can index all the state properly. Let $p^{n}_{k,j}$ be the $n$ step transition probability from the state $k$ to the state $j$.
To obtain this probabilities, we first label an index ordering by counter-clockwise from the rumor center $X_0=0$. Then, we have $P(X_{n+1} =k|X_n =k)=0$ for all $k$ and $n$, respectively. Furthermore, $P(X_{n+1} =j|X_n =k)=0$ for all $d(k,j)>1$ since the token is moved one-hop at one-step ($r$ querying). Then, the transition probability for the node $k$ which is not a leaf node in $V_L$ is as follows.
\begin{align*}
p_{k,j}=\begin{cases}
\frac{\hat{p}}{d}+(1-\hat{p})\frac{1-\hat{q}}{d-1}&\mbox{if}~j\notin nb(k,v_1)\\
\frac{\hat{p}}{d}+(1-\hat{p})\hat{q}&\mbox{if}~j \in nb(k,v_1),
\end{cases}
\end{align*}
where $nb(k,v_1)$ is the set of neighbors of the node $k$ on the path between the node $k$ and $v_1$.
%
From the simple Markov property of querying scheme, if we assume that the source node is an absorbing state then
we obtain for a given budget $K/r\geq l$,
    \begin{align}\label{eqn:transition}
     \prob{&Q_K (v_1)\geq1|d(v_1, v_{RC})=i}\cr
     &=1-\prob{Q_K (v_1)=0|d(v_1, v_{RC})=i}\cr
     &=1-\prob{\sum_{n=0}^{K}I_n (v_1)= 0|d(v_1, v_{RC})=i}\cr
     &=1-\prob{I_n (v_1)= 0,~\text{$\forall i\leq n \leq K/r$}|d(v_1, v_{RC})=i}\cr
     &=1-\prod_{n=i}^{K/r}(1-p^{n}_{0,v})\cr
     &\stackrel{(a)}{\geq}1-\prod_{n=i}^{K/r}(1-p^{K/r}_{0,v})=1-(1-p^{K/r}_{0,v})^{K/r-i}\cr
     &\stackrel{(b)}{\geq}1-e^{-(K/r -i)p^{K/r}_{0,v}},
    \end{align}
where $(a)$ follows from the fact that $p^{n}_{0,v}\geq p^{K/r}_{0,v}$ for all $i\leq n \leq K/r$ and $(b)$ is from the relation of $(1-p)^{K/r} =e^{K/r \log (1-p)}\leq e^{-p(K/r)}$ where we use the inequality $\log (1-x) \leq -x$ for $0\leq x \leq 1$. Note that the transition probability is the case of $d(v_1, v_{RC})=i$. From Lemma, we have
\begin{equation*}
    \begin{aligned}
     \prob{Q_K (v_1)\geq1}&\geq\sum_{i=1}^{\infty}(1-e^{-(K/r -i)p^{K/r}_{0,v}})P(d(v_1, v_{RC})=i)\\
     &\geq\sum_{i=1}^{\infty}(1-e^{-(K/r -i)p^{K/r}_{0,v}})\left(\frac{d-1}{d}\right)^{i} e^{-(i+1)}\\
    &\geq 1-ce^{-\frac{Kp^{K/r}_{0,v}}{r} }\geq 1-ce^{-\frac{Kp}{r}  (q-1/d)^3}.\\
    \end{aligned}
  \end{equation*}
Using this result and Lemma~\ref{lem:majority}, we conclude the result of
Lemma~\ref{lem:idinter} and this completes the proof.
\end{proof}

\smallskip
Next, we have the following result.

\smallskip
\begin{lemma}
When $p>1/2$ and $q>1/d,$
\begin{equation*}
 \begin{aligned}
\prob{v_1 \in S_D |v_1 \in V_{L}} \geq 1- e^{-\frac{Kp (d-1)(q-1/d)^2}{6rd}}
   \end{aligned}
   \end{equation*}
\label{lem:directionInter}
\end{lemma}
\smallskip

Similar to the previous one, to obtain the detection probability, we need to find the probability $P(v_1 \in S_N \cap S_I|v_1 \in V_{L})$.
From this, we consider $r$ repetition count for identity question and $r-X_{r}(v)$ for direction question where $(X_{r}(v)$
be the number of yes answers of the queried node $v$ with probability $p_v$. Hence, we see that the number of repetition count
for the direction questions is also a random variable which follows a binomial distribution with parameter $p_v$.
By considering this, We have the following lemma.

\smallskip
\begin{lemma}
Suppose $v_1 \in V_{L}$ then we have
\begin{equation}\label{eqn:candidate}
    \begin{aligned}
      P(v_1 \in S_I \cap S_D |v_1 \in V_{L})\geq 1-e^{-6g(p,q)^2 (K/r)\log r},
    \end{aligned}
  \end{equation}
where $g (p,q)=\frac{2d}{d-1} (p-1/2)^2 + \frac{d-1}{d-2} (q-1/d)^3$.
\label{lem:filter_Inter}
\end{lemma}
\smallskip

\begin{proof}
Since the events $v_1 \in S_I$ and $v_1 \in S_D$ are independent for a given $v_1 \in V_{L}$,
by using lemma \ref{lem:directionInter} and \ref{lem:directionInter} and some algebra, we have
\begin{equation*}
    \begin{aligned}
      &\prob{v_1 \in S_I \cap S_D |v_1 \in V_{L}}\\
      &\geq \left(p+(1-p)(1-e^{-(p-1/2)^{2}\log r})\right) \left(1-ce^{-\frac{Kp}{r}  (q-1/d)^3}\right)\\
      &\qquad \cdot \left(1- e^{-\frac{K (d-1)(q-1/d)^2}{6rd}}\right)\\
      &\geq  1-e^{-6g(p,q)^2 (K/r)\log r},
    \end{aligned}
  \end{equation*}
where $g (p,q)=c_1 (p-1/2)^2 + c_2 (q-1/d)^3$ for some constants $c_1$ and $c_2$ which are only depends on the degree $d$. This completes the proof of Lemma~\ref{lem:filter_Inter}.
\end{proof}

Next, we consider the following lemma which indicates the lower bound of detection probability
among the final candidate set.

\smallskip
\begin{lemma} \label{lem:filtration1}
  When $d\geq3$, $p>1/2$ and $q>1/d$,
  \begin{eqnarray*}
    \prob{v_1 =v_{LRC}|v_1 \in S_I \cap S_D}  & \geq & 1-e^{-g(p,q) K\log r}.
  \end{eqnarray*}
\end{lemma}
\smallskip

The proof technique is similar to the Lemma \ref{lem:filtration} so we omit it.
Using the obtained lemmas 5-8, we finally get the lower bound of detection probability
 of {\bf MVAD}$(r)$ for a given repetition count $r$ and this completes the proof of
 Lemma~\ref{lem:inter_prob}.
\end{proof}

The term $g_d (r,q)$ in
\eqref{eqn:detect2} is the probability that the
respondent reveals the true parent for given $r$ and $q$.
 Hence, one can see
 that for a fixed $K$, large $r$ leads to the increasing this
 probability due to the improvement for the quality of the direction answer.
 However, increasing $r$
 negatively affects the term $K/(r+1)$ in \eqref{eqn:detect2}, so that
 there is a trade off in selecting a proper $r$.
By considering the error probabilities, we obtain
\begin{equation}\label{eqn:error}
\begin{aligned}
\prob{\hat{v}\neq v_{1}}&\leq
c \cdot e^{- (K/r)\log (K/r)}+e^{-3g(p,q)^2 (K/r)\log r}\\
&+e^{-g(p,q) K \log r}\\
&\stackrel{(a)}{\leq}( c
+1)e^{-2g(p,q)^2 (K/r)(\log K/r)},
\end{aligned}
\end{equation}
where $c_1=c+1$ and $g (p,q)=\frac{2d}{d-1} (p-1/2)^2 + \frac{d-1}{d-2} (q-1/d)^3$. The inequality $(a)$ is from the fact that
$g(p,q) <1$.
By derivation of the result with respect to $r$, we first obtain $r^\star$ which maximizes the detection
probability by $r^\star= \left\lfloor1+\frac{7d^2 \{2(1-p)^3 + (1-q)^2\} \log\log
  K}{3(d-1)}\right\rfloor$ in {\bf MVAD}$(r^\star)$
and put this into the error probability $\prob{\hat{v}\neq v_{1}}$, we have
\begin{align}\label{eqn:choi}
\prob{\hat{v}\neq v_{1}}&\leq
( c
+1)e^{-2g(p,q)^2
  (K/(r^*))\log(K/(r^* ))}\cr
   &\stackrel{(a)}{\leq} ( c
+1) e^{-g (p,q)K\log (\log K)},
\end{align}
where the inequality $(a)$
comes from the obtained result of $r^\star$. Let
$\delta \geq ( c
+1) e^{-g(p,q)K\log (\log K)},$ then, we obtain the value of $K$ which
 produces the error probability $\delta$ in later and we obtain the
desired lower-bound of $K$ as in the theorem statement. This completes
the proof of Theorem~\ref{theorem:interactive}.

\vspace{-0.1cm}
\section{Simulation Results}
\label{sec:numerical}

\begin{figure}[t!]
\subfigure[Regular tree ($d=3$).]{\includegraphics[width=0.5\columnwidth]{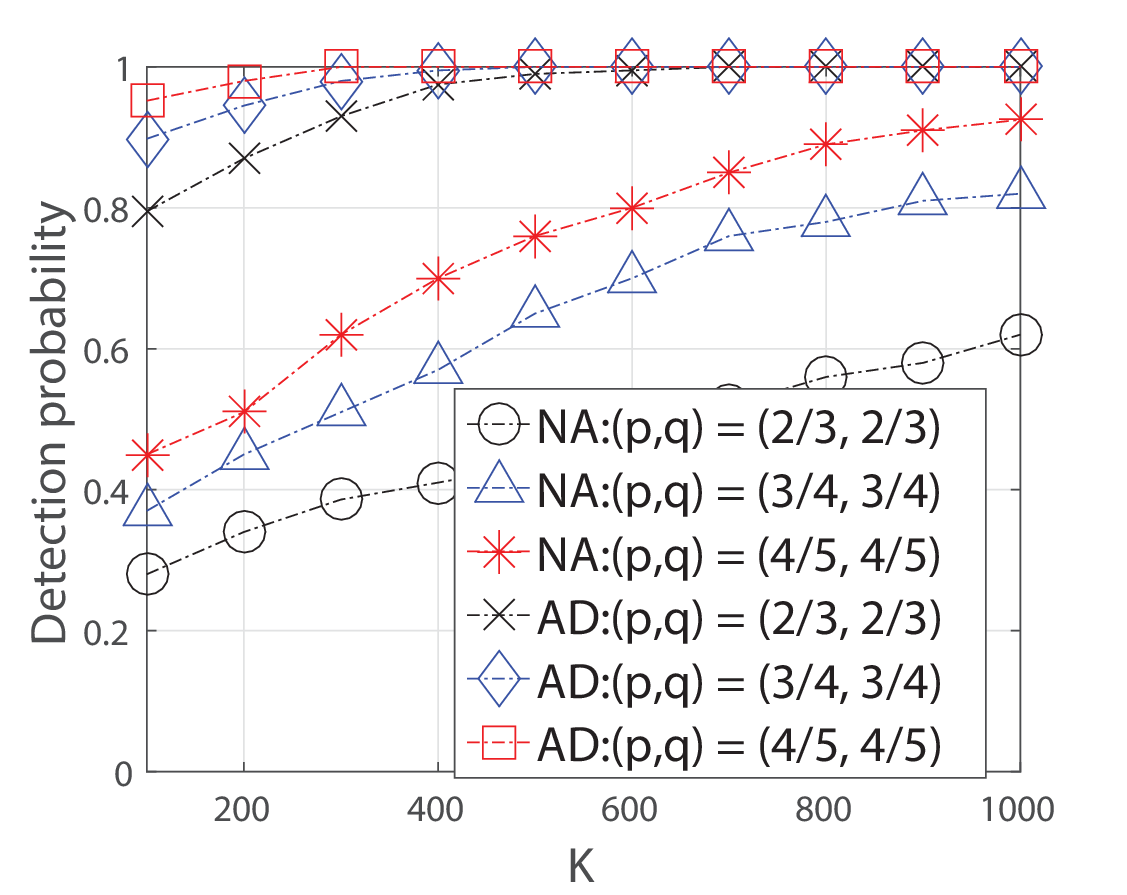}\label{fig:RGAD}}
\hspace{-0.19cm}
\subfigure[Facebook network.]{\includegraphics[width=0.5\columnwidth]{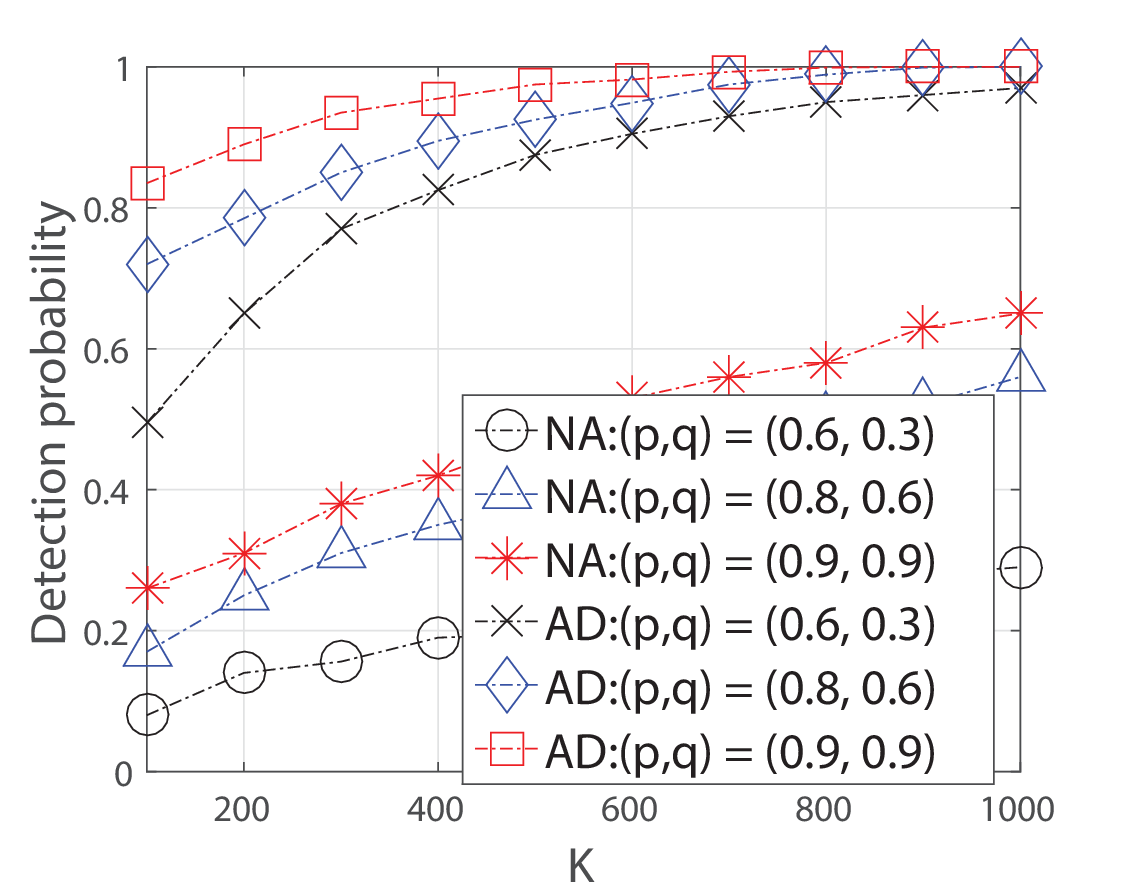}\label{fig:FBAD}}
\caption{Detection probabilities with varying $K$ for regular tree (a) and Facebook network (b), respectively.}
\label{fig:general}
\vspace{-0.3cm}
\end{figure}

In the simulation, we consider two graph topologies: regular trees,
and a Facebook graph. 
We propagate an information from
a randomly chosen node to 400 infected nodes at maximum, and plot the
detection probability from 200 iterations.
We obtain the detection probabilities with varying budgets $K$ under different parameters $(p,q).$ 
In the regular tree, we use {\bf MVNA}$(r^\star)$ and {\bf MVAD}$(r^\star)$ for both querying schemes with $d=3$ and Fig.~\ref{fig:general}(a) shows that there is a significant adaptive gain for various parameters $(p,q),$
validating our theoretical results.
Different from the regular tree, there exist loops in a general graph such as
Facebook network. It is known that computing the MLE in such a general loopy graph is \#P-complete \cite{shah2012}. Hence, as a heuristic, we use a Breath First Search (BFS) to the graph and use the BFS estimator as the initial center node of candidate set in {\bf NA}-querying and the initial node in {\bf AD}-querying, respectively.
Further, in {\bf NA}-querying, we count the number of descendants of each node in the candidate set on the BFS tree due to the loop in the general graph.
Fig.~\ref{fig:general}(b)
shows the detection probabilities with varying $K$ for {\bf NA}-querying and
{\bf AD}-querying with different parameters $(p,q)$ and we observe similar trends to those in the regular tree.
We see that the {\bf AD}-querying is powerful for finding 
the source because it uses the sampled data more efficiently in an interactive manner.

\section{Conclusion}
\label{sec:conclusion}
In this paper, we considered querying for the information source inference problem in both non-adaptive and adaptive setting. 
We obtained the answer for the fundamental
question of how much benefit adaptiveness in querying provides in finding the
source with analytical characterization in presence of individuals' untruthfulness. 




\balance
{
\renewcommand{\baselinestretch}{0.9}
\bibliographystyle{IEEEtran}
\bibliography{reference}
}

\end{document}